\documentclass{aa}
\usepackage{graphicx}
\usepackage[varg]{txfonts}
\usepackage{natbib}
\usepackage{hyperref}
\bibpunct{(}{)}{;}{a}{}{,}

\begin{document}
\title{Detecting hot stars in the Galactic centre with combined near- and mid-infrared photometry}
\author{
    M. Cano-Gonz\'alez\inst{1}
    \and 
    R. Sch\"odel\inst{2} 
    \and 
    F. Nogueras-Lara\inst{3}
    }
\institute{Universidad de Granada, Avenida de la Fuente Nueva S/N, 18071 Granada, Spain
\and Instituto de Astrof\'isica de Andaluc\'ia (IAA-CSIC), Glorieta de la Astronom\'ia s/n, 18008 Granada, Spain
\and  Max-Planck Institute for Astronomy, K\"onigstuhl 17, 69117 Heidelberg, Germany}
\date{}

\abstract {The Galactic centre (GC) is a unique astrophysical laboratory to study the stellar population of galactic nuclei because it is the only galactic nucleus whose stars can be resolved down to milliparsec scales. However, the extreme and spatially highly variable interstellar extinction towards the GC poses a serious obstacle to photometric stellar classification.} {Our goal is to identify  hot, massive stars in the nuclear stellar disc (NSD) region through combining near-infrared (NIR) and mid-infrared (MIR) photometry, and thus to demonstrate the feasibility of this technique, which may gain great importance with the arrival of the \textit{James Webb Space Telescope} (JWST).} {We combined the GALACTICNUCLEUS NIR survey with the IRAC/\textit{Spitzer} MIR survey of the GC. We applied the so-called Rayleigh-Jeans colour excess (RJCE) de-reddening method to our combined NIR-MIR data to identify potential hot stars in colour-magnitude diagrams (CMDs).} {Despite the very low angular resolution of IRAC we find 12 clear candidates for young massive stars among the $1\,065$ sources that meet our selection criteria. Seven out of these 12 stars are previously known hot, massive stars belonging to the Arches and Quintuplet clusters, as well as sources detected by the \textit{Hubble Space Telescope}/NICMOS Paschen-$\alpha$ survey. Five of our massive star candidates have not been previously reported in the literature.} {We show that the RJCE method is a valuable tool to identify hot stars in the GC using photometry alone. Upcoming instruments with high angular resolution MIR imaging capabilities such as the JWST could surely make more substantial use of this de-reddening method and help establish a far more complete census of hot, young stars in the GC area than what is possible at the moment.}
\maketitle

\section{Introduction}

The Galactic centre (GC) is a fundamental astrophysical environment to study the astrophysics of the nuclear environment of large spiral galaxies because the nearness of the GC provides us unrivalled insight because it enables us to study stars individually.

The GC region presents the most extreme conditions in the Galaxy. Stellar crowding reaches $10^{7}$ pc$^{-3}$ in the inner $0\farcs1$ of the nuclear star cluster (NSC), the massive ($2.5\pm0.4\times10^7\ M_\sun$) compact ($R_{hl}=4.2\pm0.4$ pc) cluster harbouring the $\approx4\times10^6\ M_\sun$ massive black hole Sagittarius A* (Sgr A*) located at the dynamical centre of the Galaxy \citep{Ghez2008,Genzel2010,Schodel2014,Schodel2018}. Crowding is less extreme ($\sim10^{5}$ pc$^{-3}$) in the nuclear stellar disc (NSD), a compact, disc-like structure of radius $\sim160$--$230$ pc and height of $\sim45$ pc that is often also referred to as the nuclear bulge \citep{Launhardt2002,Nishiyama2013,Gallego-Cano2020}. 

One of the many reasons to study the GC is that it has been a prolific, massive star forming region in the past few tens of millions of years, as is witnessed by the presence of three of the most massive clusters of the Galaxy (the central, Arches and Quintuplet star clusters) and numerous apparently isolated massive hot stars distributed throughout the nuclear bulge, and by its star formation history \citep[e.g.\ ][]{Figer2004,Mauerhan2010,Dong2011,NL2020_Nature}. The detection of classical Cepheids shows that about 25 Myr ago, lasting until recently, the GC underwent an increase in star formation rate relative to the previous $40$ Myr, forming around $10^6\ M_\odot$ of stars \citep{Matsunaga2011}. However, less than $10\%$ of this mass can be attributed to the three known young clusters. This may be a result of massive clusters dissolving into the crowded background within a few million years \citep{Zwart2002}.Alternatively, massive stars may form in isolation at the GC, contrary to what we find in other parts of the Milky Way. Therefore, more massive stars or clusters are expected to reside in the GC, but have not been identified yet.  The conditions in the GC are a proxy to what we can find in starburst galaxies and there is also good reason to believe that the initial mass function is top-heavy (or bottom light) at the GC \citep[see, e.g.,][]{Lu2013,Bartko2010,Hosek2019}. Hence, there is great interest to obtain a complete census of massive, hot, young stars in the GC. 

Observations of the GC suffer from ($A_V\sim30$ \,mag) extinction \citep[see][and references therein]{Nishiyama2009,Fritz2011,GALACTICNUCLEUS_III}, making it only observable at infrared (IR) wavelengths. An additional complication is that extinction varies significantly on arcsecond scales \citep[e.g.,\,][]{Scoville2003,Schodel2010,Hosek2019,GALACTICNUCLEUS_I}.
   The spectroscopic identification of stars is hampered by the extreme source crowding and by massive observational overheads. Other ways to identify massive stars are, for example, narrow band photometry \citep{Dong2011} or proper motion studies \citep{Shahzamanian2019}. Being able to use sensitive broad-band photometry from large surveys could provide significant progress.

Most of observations of the  GC stellar population are carried out in the near-infrared (NIR) regime, in particular in the $J$ ($\sim1.25\ \mu \mathrm{m}$), $H$ ($\sim1.65\ \mu \mathrm{m}$), and $K_S$ ($\sim2.15\ \mu \mathrm{m}$) bands. Unfortunately, it is very difficult to unambiguously identify hot stars in colour-magnitude diagrams (CMDs) or colour-colour diagrams (CCDs) using NIR photometry alone because of the degeneracy between reddening and stellar type \citep[see Figs.\,1 and 4 in][]{Majewski2011}. Mid-infrared (MIR) photometry can be used to break this degeneracy. Certain NIR-MIR colour combinations present little variation for a wide range of stellar types \cite[from A0 to M, see][]{Majewski2011}. 

If the intrinsic colour of a star is known, then its colour excess (the difference between the observed colour and the intrinsic colour) can be computed once an extinction law has been assumed. In particular, we can use the following expressions:

\begin{equation}\label{eq_A_Ks}
A_{K_S}= \frac{E([\mathrm{NIR}]-[\mathrm{MIR}])}{\left( \frac{\lambda_{\mathrm{NIR}}}{\lambda_{K_S}} \right)^{-\alpha_{\mathrm{NIR}}} - \left( \frac{A_{\mathrm{MIR}}}{A_{K_S}} \right)}
\end{equation}
and
\begin{equation}\label{eq_mag_0}
\mathrm{mag}(\lambda)_0 = \mathrm{mag}(\lambda)_{\mathrm{obs}}-\left(\frac{A_\lambda}{A_{K_S}}\right)A_{K_S}
\end{equation}

to calculate extinction in the $K_S$ band for  combined NIR-MIR photometry and thus retrieve  magnitudes  and colours. In these equations, $E([\mathrm{NIR}]-[\mathrm{MIR}])$ denotes the colour excess, $\alpha_{\mathrm{NIR}}$ represents the extinction index in the NIR  \citep[assuming an extinction law of the form: $A_\lambda\propto \lambda^{-\alpha}$, see][]{Nishiyama2009,Fritz2011,NL_NIR_extinction}, and $A_\lambda$ denotes extinction at a certain wavelength, $\lambda$. An extensive description of the so-called Rayleigh-Jeans colour excess (RJCE) de-reddening method can be found in \citet{Majewski2011}. Briefly stated, this method relies on the fact that intrinsic stellar colours of a broad range of stars are small and almost constant for suitable combinations of NIR and MIR filters. In particular, \citet{Majewski2011} demonstrated the usefulness of the filter combinations of $(H-[4.5\mu])$ and $(K_S-[3.6\mu])$.

 In this work we explore the application of this method to the stellar population of the GC. 
This paper is organised as follows: Section \ref{sect_data} offers a brief description of the NIR and MIR data used. In Sect. \ref{sect_metodology} we provide details on source selection and other methodological aspects of our work. In Sect. \ref{sect_results} we present the main results and in Sect. \ref{sect_conclusions}  the conclusions of our study.

\section{Data}\label{sect_data}

Our NIR data correspond to the central field of the high-resolution ($0\farcs2$) $J$, $H$, and $K_S$ GALACTICNUCLEUS  survey \citep{GALACTICNUCLEUS_II}. Its high resolution is achieved after applying the holographic imaging technique described in \cite{Schodel2013}. Details regarding data reduction and survey performance can be found in \cite{GALACTICNUCLEUS_I}.  Statistical uncertainties are below $0.05$ mag for $J\sim21\,\mathrm{mag}$, $H\sim19$ mag and $K_S\sim18$ mag and the zero-point uncertainty is $\lesssim0.04$ mag for all bands. The GALACTICNUCLEUS survey suffers from non-linearity and saturation effects in sources brighter than $K_s\lesssim11.5$ mag \citep{GALACTICNUCLEUS_II}.

For MIR photometry we used the IRAC/\textit{Spitzer} Galactic Center survey \citep{Stolovy2006}. This survey has a considerably lower resolution than GALACTICNUCLEUS with mean full width at half maximum (FWHM) of 1\farcs66, 1\farcs72, 1\farcs88, and 1\farcs98 for channels 1, 2, 3, and 4, respectively, which correspond to broadband MIR filters approximately centred on $3.6\ \mu\mathrm{m}$, $4.5\ \mu\mathrm{m}$, $5.8\ \mu\mathrm{m}$, and $8.0\ \mu\mathrm{m}$. 
 We used the IRAC point source catalogue provided by \cite{Ramirez2008}. Its average confusion limits are 12.4, 12.1, 11.7, and 11.2 mag for channels 1 to 4. 

Data from the Two Micron All Sky Survey (2MASS) at $J$, $H$, $K_S$  are also included in the \cite{Ramirez2008} catalogue. The 2MASS survey provides point sources with $\gtrsim 2\farcs5$ mean FWHM for sources in the brightness range $(4\ \mathrm{mag}\lesssim K_S^{\mathrm{2MASS}}\lesssim14\ \mathrm{mag})$ \citep{Skrutskie2006}.

In addition, we also used  multi-epoch data from the SIRIUS/IRSF GC survey \citep{Nagayama2003,Nishiyama2006a,Hatano2013} to detect and deselect variable stars (see Sect. \ref{variability_study}). The SIRIUS GC survey is seeing-limited (FWHM: $\gtrsim1\arcsec$) with $10\sigma$ limiting magnitude averages of $J^{\mathrm{SIRIUS}}=17.1$ mag, $H^{\mathrm{SIRIUS}}=16.6\, \mathrm{mag}$, and $K_S^{\mathrm{SIRIUS}}=15.6\,\mathrm{mag}$ \citep{Nishiyama2006a}. Saturation and non-linearity effects are prominent for magnitudes brighter than $9.5$, $9.5$, and $9.0$ mag for $J^{\mathrm{SIRIUS}}$, $H^{\mathrm{SIRIUS}}$, and $K_S^{\mathrm{SIRIUS}}$ bands, respectively \citep{Matsunaga2009}.

\section{Methodology}\label{sect_metodology}
In the following subsections we describe how we combined the catalogues, how we deselected variable stars and outliers, and how we identified the hot star candidates.
\subsection{Catalogue merging}\label{merging}
\begin{figure}
\includegraphics[width=\columnwidth]{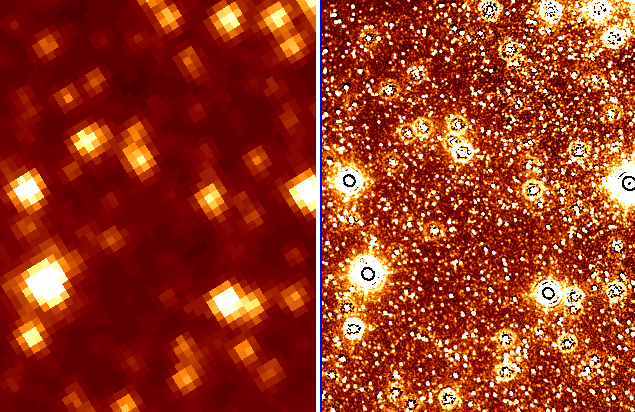}
\caption{\textit{Left:} Detail of IRAC $[3.6\mu]$ image. \textit{Right:} Same area as seen by GALACTICNUCLEUS $K_S$ band. This image illustrates the considerable difference in angular resolution between the main surveys used in this work. The dark rings present in GALACTICNUCLEUS brightest stars (also seen in Figs. \ref{Arches_det} and \ref{Quint_det}) are a consequence of a combined effect of the holographic data reduction technique and the image scaling. These stars are not marked or outlined in any way.}
\label{IRAC_GNS_res_comparison}
\end{figure}

Since there is about an order of magnitude difference in the angular resolution between the IRAC/\textit{Spitzer}, 2MASS, SIRIUS data, and GALACTICNUCLEUS data (see Fig. \ref{IRAC_GNS_res_comparison}), we applied the following merging criteria:

1) We imposed a preliminary matching radius of $0\farcs265$. 2) We determined the mean coordinate offsets between the data and corrected them. Subsequently, we repeated cross-identification with a smaller matching radius of $0\farcs133$. 3) Some apparently single IRAC/2MASS/SIRIUS sources have several NIR counterparts in GALACTICNUCLEUS because of the higher resolution of the latter survey. Therefore, we deselected sources with close neighbours that may contaminate the IRAC photometry. We excluded stars that had another GALACTICNUCLEUS detection within the IRAC channel 2 mean half width at half maximum (HWHM) ($0\farcs86$; \citealt{Fazio2004}).

With these criteria we obtained a total of $5\,585$ matching NIR-MIR detections.

\subsection{Variability}\label{variability_study}
All data used in this work were taken at different epochs. The 2MASS survey collected data between 1997 and 2001, \textit{Spitzer} performed its GC survey with IRAC in 2005, and GALACTICNUCLEUS started in 2015. These differences in observing epochs result in unphysical observed stellar colours for many variable stars. To account for variability, we used two SIRIUS/IRSF NIR data sets \citep{Nishiyama2006a,Hatano2013} as well as GALACTICNUCLEUS and 2MASS data from \cite{Ramirez2008} for bright, saturated sources in GALACTICNUCLEUS. We took the following steps: 1) We used the SIRIUS survey as a photometric reference and adjusted the median magnitude offsets of GALACTICNUCLEUS, 2MASS, and the other SIRIUS data \cite[polarimetry study by][]{Hatano2013} to the SIRIUS photometry. Only bright ($K_S<13.5$\,mag), non-saturated detections were used for this purpose. 2) For each NIR band and each catalogue combination (SIRIUS-SIRIUS\textsubscript{pol}, SIRIUS-GALACTICNUCLEUS, SIRIUS\textsubscript{pol}-GALACTICNUCLEUS) we computed
\begin{equation}
\chi_{\mathrm{var}}=\frac{|\mathrm{mag_1}-\mathrm{mag_2}|}{\sqrt{\mathrm{emag_1}^2+\mathrm{emag_2}^2}}
\end{equation}
for every common detection. If this quantity was larger than three for any band and any combination of two measurements, that particular star was flagged as variable.

With this procedure, $2\,628$ stars appeared to be non-variable and $1\,907$ stars were flagged as variable for at least one catalogue combination. Therefore, $\approx42\%$ of the stars in our sample are flagged as potentially variable. Finally, $1\,050$ stars were not detected in some of the above-mentioned catalogues, making it thus impossible to infer their variability with this method. We excluded both these unclassified stars and the variable stars from our selection.

\subsection{De-reddening}\label{sect_de-reddening}
We now proceeded to de-redden the stars in our combined NIR-MIR GC catalogue of non-variable stars with the RJCE method. We computed extinction in the $K_S$ band and intrinsic magnitudes and colours with Eqs.\,(\ref{eq_A_Ks}) and (\ref{eq_mag_0}). We used the filter combinations with the narrowest distribution of intrinsic colours according to \citet{Majewski2011}: $(H-[4.5\mu])$ and $(K_S-[3.6\mu])$.  We assumed intrinsic colour values of $(H-[4.5\mu])_0=0.08\,\mathrm{mag}$  and $(K_S-[3.6\mu])_0=0.08\,\mathrm{mag}$ \citep[based on Fig.\,3 in][and on our own estimates from stellar atmosphere models]{Majewski2011}.

\begin{figure}
\includegraphics[width=\columnwidth]{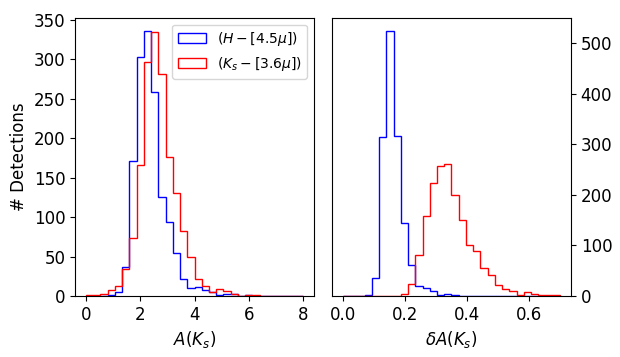}
\caption{\textit{Left:} $A(K_S)$ histograms for the combined NIR-MIR sources that met the photometric quality criteria and the foreground cut: $(H-[4.5\mu])$ (blue) and $(K_S-[3.6\mu])$ (red). \textit{Right:} Corresponding distributions of the uncertainties for the same stars and colours. The considerably higher uncertainties of $(K_S-[3.6\mu])$ are noted.}
\label{A_Ks_histograms}
\end{figure}

The resulting  distributions of $A_{K_S}$ extinction and its corresponding uncertainties (based on the photometric uncertainties) for the two colours are shown in Fig.\,\ref{A_Ks_histograms}. We found that $(H-[4.5\mu])$ provided more reliable results in terms of de-reddened colours and uncertainties than $(K_S-[3.6\mu])$. This could be explained by the larger wavelength range covered by the former, hence reducing systematics related to filter width, and the deeper dust penetration capabilities of the latter. For this reason we only use $(H-[4.5\mu])$ henceforth. We only used stars with with uncertainties $<0.1$ mag in both GALACTICNUCLEUS and IRAC photometry. We excluded foreground stars belonging to the Galactic disc and bulge using a $(H-K_S)<1.3\, \mathrm{mag}$ colour cut \citep{NL2020_Nature,NL_2021}. For the extinction curve, we adopted  values of  $\alpha_{\mathrm{NIR}}=2.11\pm0.06$ and $A_{[4.5\mu]}/A_{K_S}=0.396\pm0.082$, $A_{[3.6\mu]}/A_{K_S}=0.547\pm0.052$ from \cite{Fritz2011}.

\section{Results and discussion}\label{sect_results}
\begin{figure}
\includegraphics[width=\columnwidth]{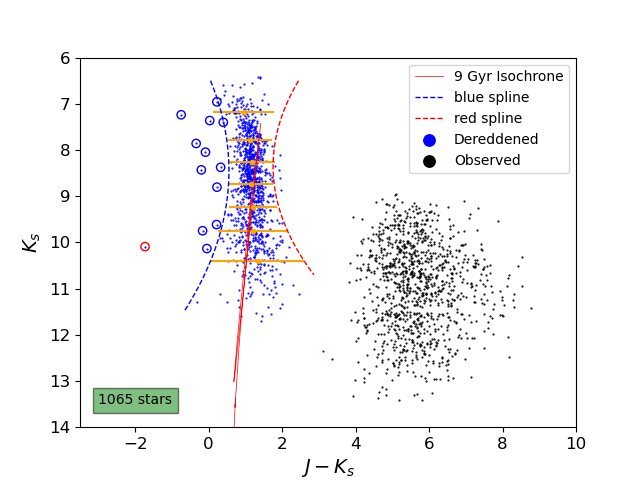}
\caption{Final $K_S$ vs $J-K_S$ CMD. Both observed (black) and de-reddened (blue) data are plotted. We fitted the de-reddened RGB cloud to a 9 Gyr, $Z=0.039$ isochrone (red line). We separate $12$ hot stars (blue circles) from the RGB. }
\label{final_CMD}
\end{figure}

The $(J-K_S,K_S)$ CMD of our final sample  is shown in Fig.\,\ref{final_CMD}. We chose this colour combination because the $H$ band was already used in the de-reddening process, and because the intrinsic colour of stars is expected to be large for this filter combination

All stars plotted in Fig. \ref{final_CMD} are detected in all three GALACTICNUCLEUS bands and in the  IRAC channel\,2 ($4.5\ \mu$m), meet the non-variability criteria, have photometric uncertainties below $0.1$ mag, and are not foreground sources. Additionally, we excluded faint $(K_S>13.5\ \mathrm{mag})$ sources because their de-reddened counterparts presented extremely blue ($(J-K_S)_0\approx-2,-3$ mag) colours.  We suppose that this over-correction was related to systematic errors of the IRAC photometry of such faint sources in this extremely crowded field.

We  fitted by eye a 9 Gyr, $Z=0.039$, solar-scaled BaSTI isochrone model to the main cloud of points in the CMD \citep{BaSTI,NL2020_Nature}. For this purpose we used $\alpha_{\mathrm{NIR}}=2.11$ \citep{Fritz2011} and the ratio $A[4.5\mu]/A_{K_S} = 0.32$. The latter ratio is lower than -- but within the uncertainties provided for this ratio by \citet{Fritz2011}. Small changes in the extinction index and/or ratios translate to small variations in the linear dependence of extinction with respect to colour excess, that is $A(K_S)=\varphi\ E(H-[4.5\mu])$. In other words, small changes ($A_{[4.5\mu]}/A_{K_S}=0.32$ instead of $A_{[4.5\mu]}/A_{K_S}=0.396$) in the $\varphi$ coefficient yield changes for all detections, modulated by their colour excess. This effect is therefore of no concern  when separating hot stars in a de-reddened CMD using photometry alone, which is our main purpose rather than providing an accurate measurement of de-reddened magnitudes and colours.

We note that reasonably varying either age $(7-10\ \mathrm{Gyr})$ and/or metallicity $(0.02<Z<0.04)$ resulted in little change in the position of the isochrone in the CMD. In comparison to the isochrone, the red giant branch (RGB) appears somewhat tilted towards the blue with increasing brightness. We were not able to identify the reason for this behaviour, which may be caused by a complex relation between effects such as non-linearity of photometry in some of the data, extinction correction, and effective wavelength in broad-band filters, or the precise composition of the stellar population; the effective wavelength results from a complex function of filter, stellar spectral energy distribution (SED), extinction curve, and absolute extinction. The important observation is, however, that the RGB stars cluster along a tight sequence in the de-reddened RGB. 

We separated the de-reddened RGB cloud into approximately 0.5 magnitude bins ($(K_S)_0<7.5\,\mathrm{mag}$, $7.5\,\mathrm{mag}<(K_S)_0<8\,\mathrm{mag}$, $8\,\mathrm{mag}<(K_S)_0<8.5\,\mathrm{mag}$, $8.5\,\mathrm{mag}<(K_S)_0<9\,\mathrm{mag}$, $9\,\mathrm{mag}<(K_S)_0<9.5\,\mathrm{mag}$, $9.5\,\mathrm{mag}<(K_S)_0<10\,\mathrm{mag}$ and $10\,\mathrm{mag}<(K_S)_0$) and computed the median and standard deviation of $(J-K_S)_0$ and $(K_S)_0$ for each bin. We chose this bin width so at least 100 stars per bin are used. Each bin contains $\sim150$ stars on average.

We identified as hot, and presumably young, those stars bluer than $2.5\sigma$ from the median $(J-K_S)_0$ in each bin. As shown in Fig. \ref{final_CMD}, the de-reddened cloud widens for magnitudes brighter than $(K_S)_0\sim10\,\mathrm{mag}$. Therefore, to separate real hot detections from the blue tail of the RGB quasi-Gaussian distribution, we interpolated with a spline degree two (blue, dashed line) using as nodes those values located $2.5\sigma$ bluer than the median $(J-K_S)_0$ of each bin. The position of the nodes in the magnitude axis is given by the median $(K_S)_0$ of each bin. Thus, $12$ out of the $1\,065$ stars plotted are identified as hot (blue circles). We interpret one star, which is bluer than $7\sigma$ from the median $(J-K_S)_0$,  as an outlier (red circle). We proceeded analogously with the red tail of the de-reddened RGB. Only five sources are redder than $2.5\sigma$ from the median $(J-K_S)_0$, suggesting that our hot candidates are hot stars and not residual errors from the de-reddened distribution.

Since IRAC photometry is less affected by interstellar extinction than $K_S$-band photometry, in principle, some highly reddened sources could be detected at $[3.6\mu]$ or $[4.5\mu]$ while being fainter than the $K_S$ detection limit. This scenario could result in possible contamination due to chance alignments of MIR sources that are not detected in the NIR and therefore not deselected in the NIR-only neighbour exclusion criteria explained in Sect. \ref{merging}. To test whether this concerns our case we used average intrinsic (non-reddened) $(K_S-[4.5\mu])_0$ colours of several stellar types and reddened them with typical GC extinction values using the extinction coefficients from \citet{Fritz2011}. For RGB stars, we averaged the values given in the IRAC K-III primary calibrators \citep[][Table\,1]{Reach2005} and the intrinsic colours for M-type giants provided in \citet[][Table\, 11]{Jian2017}. Both K-type and M-type giants showed similar intrinsic NIR-MIR colours: $(K_S-[4.5\mu])_0\sim0\, \mathrm{mag}$. Therefore, given that our faintest $[4.5\mu]$ detections are $\sim12\, \mathrm{mag}$, we estimated how much fainter such detections would appear in the $K_S$ band and checked if they were above the GALACTICNUCLEUS survey $K_S$ detection limit ($K_S\sim18\, \mathrm{mag}$). In an average extinction scenario ($A(K_S)\sim 2\, \mathrm{mag}$ and $A([4.5\mu])\sim1\,\mathrm{mag}$) such stars would be clearly detectable by GALACTICNUCLEUS since they would be around $K_S\sim15\,\mathrm{mag}$: $(K_S-[4.5\mu])=(K_S-[4.5\mu])_0 + A(K_S)+A([4.5\mu])$. Even in a high extinction scenario ($A(K_S)\sim 3.5\, \mathrm{mag}$ and $A([4.5\mu])\sim1.5\,\mathrm{mag}$), such highly reddened RGB stars would present $K_S\sim 17\,\mathrm{mag}$, brighter than GALACTICNUCLEUS detection limit. We proceeded analogously with asymptotic giant branch (AGB) stars, whose intrinsic colours we derived from \citet[][Tables\, 1 and 2]{Reiter2015}. We used the five closest stars from each sub-category (O-rich, C-rich and S-rich) to minimise extinction, and averaged the values. We obtained the following results: $(K_S-[4.5\mu])_0\approx0.2\pm0.3\,\mathrm{mag},0.5\pm0.2\,\mathrm{mag}$ and $0.3\pm0.3\,\mathrm{mag}$ for O-rich, C-rich, and S-rich AGB stars respectively. We note that using the multi-epoch $[4.5\mu]$ values given in \citet[][Table\,  2]{Reiter2015} did not change our average results. Thus, even if we assume the redder colours from C-rich AGB stars and a high extinction scenario, the $K_S$-band counterpart of such stars would be around $K_S\sim17.5\,\mathrm{mag}$ and therefore detectable by GALACTICNUCLEUS. Regarding the possibility of contamination due to objects with intrinsically redder SEDs such as young stellar objects (YSOs) we rely on the results from \cite{Nandakumar2018}. This spectroscopic study showed that many of the objects previously identified as YSOs with CMD were cool, evolved AGB stars, indicating how similar NIR and MIR colours are for AGB stars and YSOs. Furthermore, the total amount of YSOs detected in the central molecular zone (CMZ) is relatively small (a few hundreds, according to \citealt{Yusef-Zadeh2009}); therefore, the chance alignment of such an object within $\sim1 \arcsec$ of one of our target stars is very small.

\begin{figure*}
\centering
\includegraphics[width=\textwidth]{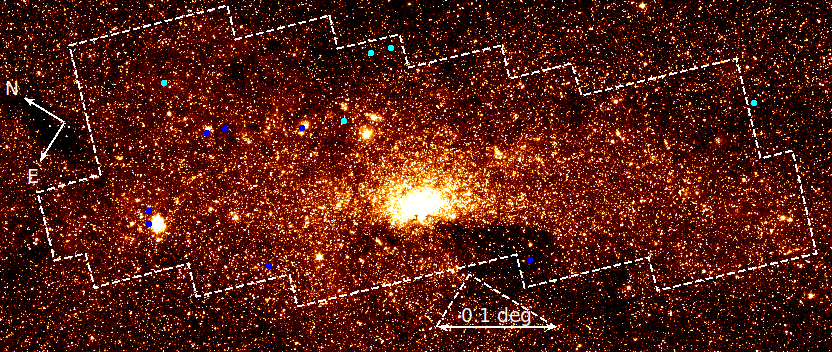}
\caption{Hot star candidates shown over the IRAC $[4.5\mu]$ image. The image is approximately centred on the Milky Way NSC. The HST NICMOS Paschen-$\alpha$ survey area is indicated by the white dashed lines. The dark blue circles show previously known massive stars belonging to our candidate list. The light blue circles show the positions of our hot candidates that have not been reported in previous literature.}
\label{hot_stars_field}
\end{figure*}

We used the NICMOS HST Paschen-$\alpha$ survey \citep{Wang2010,Dong2011} to test whether we could find any previously detected massive stars. The majority of hot stars reported by \citet{Dong2011} are not present in our final data set because of our selection criteria. However, we find seven hot massive stars from their primary Paschen-$\alpha$ emitting candidates \citep[][Table 3]{Dong2011}. We find none of their secondary emitting Paschen-$\alpha$ sources. Our hot stars candidates, and their photometric properties are listed in Table \ref{table_hs} and their positions on the sky are represented in Fig. \ref{hot_stars_field}.

\begin{table*}
\caption{Hot star candidates. J2000 right ascension (R.A.) and declination ($\delta$) taken from GALACTICNUCLEUS. All \textbf{NIR} measurements are those of GALACTICNUCLEUS, except for stars brighter than $K_S=11.5$ mag whose $K_S$ photometry is taken from 2MASS. The IRAC MIR photometry is that of \cite{Ramirez2008} catalogue. The last column shows the ID of previously known hot stars: $a$: \citet[Table\, 3]{Dong2011}; $b$: \cite{Clark2018_Arches}; and $c$: \cite{Clark2018}. The label \texttt{NaN} means no detection in a particular band.}             
\label{table_hs}      
\centering                          
\begin{tabular}{c c c c c c c c c c c c c}        
\hline\hline                 
R.A. (deg) & $\delta$ (deg) & $J$ & e$J$ & $H$ & e$H$ & $K_S$ & e$K_S$ &  $[3.6\mu]$ & e$[3.6\mu]$ &  $[4.5\mu]$ & e$[4.5\mu]$ &   ID\\    
\hline                        
$266.38721$ & $-28.91581$ & $18.57$ & $0.10$ & $14.57$ & $0.08$ & $12.66$ & $0.09$ & \texttt{NaN} & \texttt{NaN} & $10.16$ & $0.01$ &    \\ 
$266.16985$ & $-29.19812$ & $15.85$ & $0.09$ & $11.95$ & $0.09$ & $9.77$ & $0.03$ & $7.89$ & $0.01$ & $7.69$ & $0.01$ &    \\ 
$266.31870$ & $-28.90582$ & $20.20$ & $0.10$ & $14.64$ & $0.09$ & $11.42$ & $0.05$ & $9.01$ & $0.01$ & $8.57$ & $0.01$ &    \\ 
$266.30472$ & $-28.91789$ & $17.41$ & $0.09$ & $13.12$ & $0.09$ & $10.78$ & $0.02$ & $8.64$ & $0.01$ & $7.77$ & $0.01$ &    \\ 
$266.45230$ & $-28.83499$ & $17.29$ & $0.09$ & $13.27$ & $0.09$ & $11.02$ & $0.05$ & $9.10$ & $0.01$ & $8.48$ & $0.01$ &   $96^a$ \\ 
$266.55600$ & $-28.81649$ & $14.58$ & $0.10$ & $11.58$ & $0.10$ & $9.69$ & $0.03$ & $8.45$ & $0.01$ & $8.03$ & $0.01$ &   $4^a,31^c$ \\ 
$266.54169$ & $-28.92569$ & $15.03$ & $0.08$ & $12.36$ & $0.10$ & $10.75$ & $0.04$ & $9.72$ & $0.01$ & $9.38$ & $0.01$ &   $23^a$ \\ 
$266.41382$ & $-28.88921$ & $14.57$ & $0.09$ & $11.79$ & $0.09$ & $10.19$ & $0.05$ & $8.94$ & $0.01$ & $8.52$ & $0.01$ &   $106^a$ \\ 
$266.40787$ & $-29.10816$ & $15.74$ & $0.09$ & $13.03$ & $0.09$ & $11.53$ & $0.02$ & $10.54$ & $0.01$ & $10.09$ & $0.01$ &   $40^a$ \\ 
$266.46439$ & $-28.82381$ & $16.78$ & $0.08$ & $13.19$ & $0.08$ & $11.20$ & $0.07$ & $9.72$ & $0.01$ & $8.99$ & $0.01$ &   $78^a,1^b$ \\ 
$266.44547$ & $-28.77219$ & $17.81$ & $0.09$ & $14.37$ & $0.08$ & $12.63$ & $0.09$ & $10.86$ & $0.02$ & $10.56$ & $0.02$ &    \\ 
$266.56671$ & $-28.82275$ & $15.18$ & $0.09$ & $12.24$ & $0.09$ & $10.47$ & $0.05$ & $9.37$ & $0.05$ & $9.03$ & $0.06$ &   $72^a,21^c$ \\ 

\hline                                   
\end{tabular}
\end{table*}

\begin{figure*}
\includegraphics[width=\textwidth]{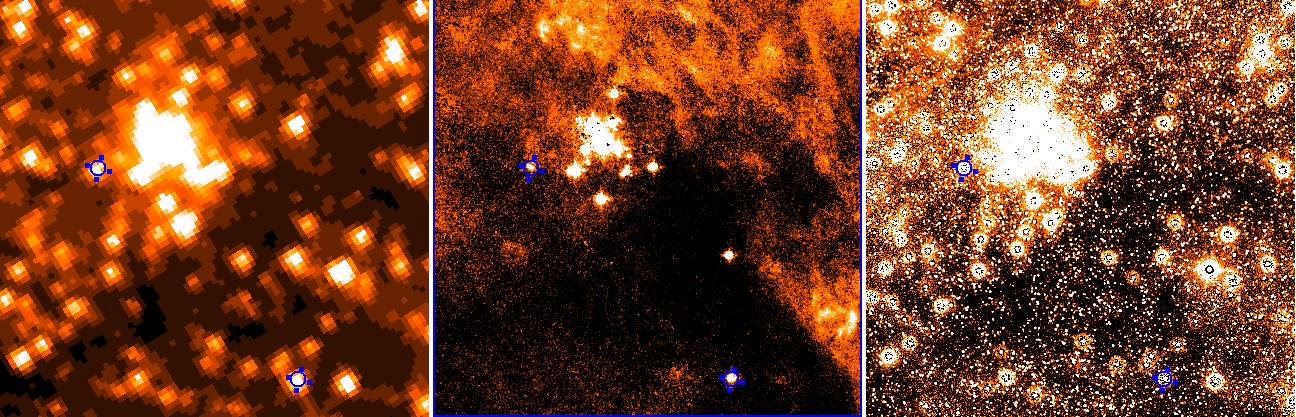}
\caption{Hot stars (blue circles) identified in the outskirts of the Arches cluster; only the top-most hot stars belongs to the cluster. From left to right: IRAC channel 2 ($[4.5\mu\mathrm{m}]$), HST NICMOS Paschen-$\alpha$ image and GALACTICNUCLEUS $K_S$ band image.}
\label{Arches_det}
\end{figure*}

\begin{figure*}
\includegraphics[width=\textwidth]{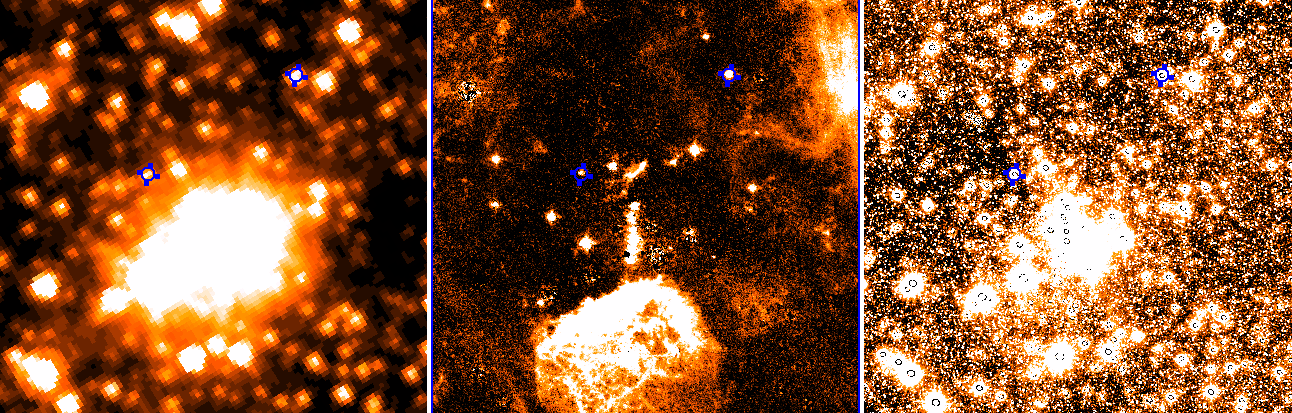}
\caption{Same as in Fig. \ref{Arches_det} but for the Quintuplet cluster.}
\label{Quint_det}
\end{figure*}

According to recent  census of massive young stars in  the Arches and Quintuplet clusters \citep{Clark2018_Arches,Clark2018}, we detect one star belonging to the former cluster (top circled detection of Fig \ref{Arches_det}) and two stars belonging to the latter cluster (both circled stars in Fig. \ref{Quint_det}).

Presumably, we do not detect more known massive stars belonging to either cluster because of our strict close neighbour exclusion, and because the IRAC high confusion limits in such crowded fields.

\section{Conclusions}\label{sect_conclusions}

Studies of star formation at the GC are limited by interstellar extinction that complicates the application of broad-band photometry for even a rough stellar classification. The RJCE method combines NIR and MIR data and uses the fact that most stars have a very narrow distribution of intrinsic colours in given combinations of NIR and MIR filter and is of great interest for highly extinguished regions of the Galaxy. Unfortunately there is no way to image large fields from the ground in the MIR, which can only be done from space. The upcoming JWST, with its high angular resolution ($\lesssim0\farcs2$ at wavelengths ($\lesssim5\,\mu$m)) can therefore provide a breakthrough in our understanding of star formation at the GC. 

Currently, we have only low angular resolution MIR photometry from IRAC/\textit{Spitzer} available. We used these data and combined them with the high angular resolution NIR GALACTICNUCLEUS survey in a pilot study to demonstrate the application of the RJCE method to the GC. Even though we are only left with about 1\,000\,stars in our sample after applying the necessary data selection criteria, we can  identify candidates for hot stars in the de-reddened CMD. The latter are placed in a separate sequence that is clearly placed to the blue of the RGB. About half of our candidates are known from previous studies and some of these are known members of the  Arches and Quintuplet clusters. We also report on five hot star candidates that have not been previously reported in the literature.

\begin{acknowledgements}
      R. S. acknowledges financial support from the State Agency for Research of the Spanish MCIU through the ``Center of Excellence Severo Ochoa'' award for the Instituto de Astrof\'isica de Andaluc\'ia (SEV-2017-0709). R. S. acknowledges financial support from national project PGC2018-095049-B-C21 (MCIU/AEI/FEDER, UE).
      
      F.N.-L. gratefully acknowledges support by the Deutsche Forschungsgemeinschaft (DFG, German Research Foundation) – Project-ID 138713538 – SFB 881 ("The Milky Way System", subproject B8).
\end{acknowledgements}

\bibliographystyle{aa}
\bibliography{RJCE_GC_draft.bib}

\end{document}